# Self-Attention and Hybrid Features for Replay and Deep-Fake Audio Detection

Lian Huang, and Chi-Man Pun

***Abstract*—Due to the successful application of deep learning, audio spoofing detection has made significant progress. Spoofed audio with speech synthesis or voice conversion can be well detected by many countermeasures. However, an automatic speaker verification system is still vulnerable to spoofing attacks such as replay or Deep-Fake audio. Deep-Fake audio means that the spoofed utterances are generated using text-to-speech (TTS) and voice conversion (VC) algorithms. Here, we propose a novel framework based on hybrid features with the self-attention mechanism. It is expected that hybrid features can be used to get more discrimination capacity. Firstly, instead of only one type of conventional feature, deep learning features and Mel-spectrogram features will be extracted by two parallel paths: convolution neural networks and a short-time Fourier transform (STFT) followed by Mel-frequency. Secondly, features will be concatenated by a max-pooling layer. Thirdly, there is a Self-attention mechanism for focusing on essential elements. Finally, ResNet and a linear layer are built to get the results. Experimental results reveal that the hybrid features, compared with conventional features, can cover more details of an utterance. We achieve the best Equal Error Rate (EER) of 9.67% in the physical access (PA) scenario and 8.94% in the Deep fake task on the ASVspoof 2021 dataset. Compared with the best baseline system, the proposed approach improves by 74.60% and 60.05%, respectively.**

## I. Introduction

VOICE assistants have been more and more widely used recently. Although they are convenient, the related security issues cannot be ignored. An audio Speaker Verification (ASV) system should be required to ensure our voice assistant is secure and not used illegally by others [1]. To address this issue, there are four challenges with public datasets from 2015 to 2021 [2]–[5]. With the outstanding efforts of community scholars, the performance of the ASV system has been significantly improved. Unfortunately, the ASV is still vulnerable to spoofing attacks in practical applications. For example, with the support of deep learning technology, the speech generated by many text-to-speech (TTS) and voice conversion (VC) algorithms is getting closer and closer to natural speech [6], [7]. It is hard to detect this deep fake speech. The countermeasure that achieved the best performance on the task named speech Deep-Fake (DF) in the ASV spoof 2021 challenge had an Equal Error Rate (EER) of 15.64% [5]. A similar situation also occurs in replay attacks [8], [9].

Lian Huang is with the Department of Applied Electronics, Guangdong Mechanical and Electrical College, Guangzhou 510006, China. E-mail: mrhuanglian@gmail.com.

Chi-Man Pun is with the Department of Computer and Information Science, University of Macau, Macao 999078, China. E-mail: cmpun@umac.mo.

Furthermore, replay attacks are easy to launch [10]. When an attacker wants to gain access to secured content by replay spoofing attacks, it does not need special signal processing knowledge but just a recording and playback device. Such as a mobile phone, the whole process of spoofing attack can be implemented quickly. Fig. 1 shows the process of replay or DF attack. In this paper, we focus on detecting DF and replay spoof audio.

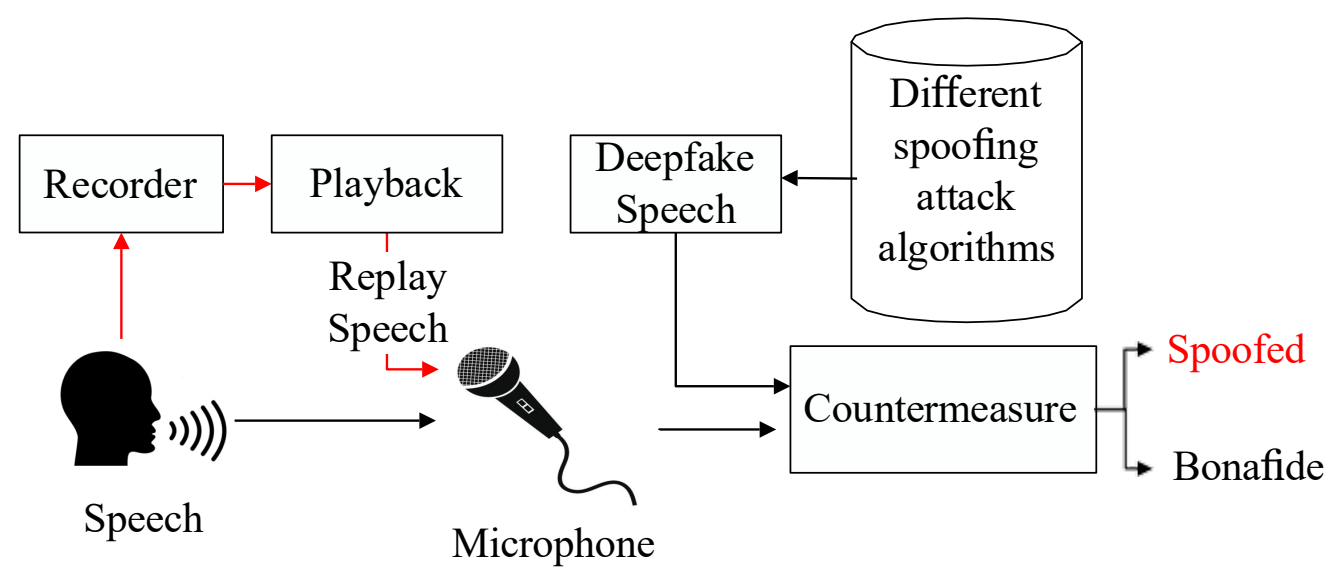

Fig. 1. Replay or Deep-Fake attack process.

Spoofed audio detection has attracted the attention of many scholars in the community and has made significant progress. The ASV spoof2015 challenge focuses on speech synthesis (SS) and voice conversion (VC) cases. There are ten different spoof types. The fusion method [11] got the best result in that challenge, but that system is too complex and less general. After that challenge, Todisco *et al.* almost got perfect performance with the CQCC features except for the unknown types spoof method [12]. In [13], Yang *et al.* proposed three subband transform features and got promising results. However, all those systems did not mention replay attack detection. The BTAS 2016 speaker anti-spoofing competition began to include replay detection tasks [14]. Then the ASVspoof2017 challenge only focused on the detection of replay speech. Detecting the audio replay spoofing attack is more challenging than the others [3]. In [15], Chettri *et al.* believed that many successful countermeasures took advantage of the fact that the database contains some recording artifacts. An utterance of replay spoofing almost contains all the features that are in the genuine one. In [8], You *et al.* took device features to detect replay audio and got an EER of 8.99% in the ASVspoof 2017 corpus version 2.0. But with high-quality equipment, we nearly cannot distinguish all utterances by our naked ears in the ASV spoof 2019 dataset in the physical access (PA) scenario, which is the replay spoof attack. In this scenario, it is hard to get the device feature to detect the replay spoof audio.

Currently, countermeasures can achieve better results under a specific dataset training [16]–[20], but it is prone to over-fitting. For example, in the ASVSpoof2019 challenge, the top 10 methods are in the 2019 real PA. Its performance has dropped significantly, even inferior to Baseline [21]. It is indeed challenging to detect replay spoofs in practical applications. Ahmed *et al.* proposed a relatively simple model in [22]. By detecting liveness, good results have been achieved. Its advantage is that the model is relatively simple and can be directly deployed to the user terminal. Its performance against high-quality speakers may degrade. In [23], Md. Sahidullah *et al.* add additional devices, such as throat microphones, to get better accuracy.

Except for the detection of logical access (LA) and PA, the detection of DF is another task in the ASVspoof2021. All DF speeches were generated using more than one hundred TTS or VC algorithms [5]. The results of the ASVspoof 2021 challenge showed that PA and DF tasks still need more solutions. The best EER is 24.25% and 15.64%, respectively. In [24], Jun *et al.* proposed the method based on the fundamental frequency information and got almost the best EER in the ASVspoof 2019 LA dataset. In [7], Nishant Subramani and Delip Rao tried to use transfer learning for the detection of fake speech. But all of them cannot reproduce the same good result in the DF task in the ASVspoof2021. Even though most of them can achieve promising performance with training or development data, worse evaluation results show that the system's generality still needs improvement.

To detect deep synthesis and manipulated audios, the 2022 Audio Deep synthesis Detection (ADD) challenge launched with three tracks: 1) Low-quality fake audio detection, 2) Partially fake audio detection, and 3) Audio fake game [25]. Track 1 is similar to the DF task in the ASVspoof2021. But there are some background noises and disturbances in this task. With unsupervised pretraining models, [26] got an EER of 4.8% in Track 2 but got a worse result with an EER of 32.8% in Track 1. DF detection is still a very tricky task.

For the detection of unseen attacks invented in the future, it still needs to improve the performance of countermeasures. Most previous work focuses on different traditional features or enhanced classifiers with deep learning frameworks. And some other methods with complex architectures can get better performance in training. But the results at the evaluation stage of these systems show more or less over-fitting. This paper proposes a novel solution, using hybrid features based on weighted fusion with the self-attention mechanism and using ResNet [27] as a back-end classifier.

The contributions of this paper can be summarized as follows:

1. We explore integrating deep learning features and Mel-spectrogram features to replace the previous traditional features.

2. We propose a self-attention mechanism after the fusion of features. The primary purpose of this mechanism is to focus on the essential elements of spoofed speech. The classifier uses ResNet for better generality.

3. Experimental results show that our method achieves the best Equal Error Rate (EER) of 9.67% in the physical access (PA) scenario and 8.94% in the Deep fake task on the ASVspoof 2021 dataset. Compared with the best baseline system, our method improves by 74.60% and 60.05%, respectively.

The rest of this paper is organized as follows. Section II introduce related work, including replay and DF speech, audio features, and some classifier in the field of spoof audio detection. Section III elaborates on our proposed method, including motivation, feature extraction, Self-attention, and classifier. The following Section IV is the verification experiment, and Section V is the discussion part. The conclusion is in Section VI.

## II. Related Work

### A. Replay and Deep-Fake Speech

A replay attack is the most accessible spoofing and is an example of a low-effort spoofing attack [14]. Compared with genuine speech, replay speech carries additional information generated by recording and playback devices. Considering the differences in environmental conditions, genuine speech signal $S_{genuine}(t)$ and replay speech signal $S_{replay}(t)$ can be defined as:

$$S_{genuine}(t) = S_h(t) * E(t) * M_{mic}(t) \quad (1)$$

$$S_{replay}(t) = S_h(t) * M'_{mic}(t) * S_{spk}(t) * E'(t) * M_{mic}(t) \quad (2)$$

where *denotes convolution, $S_h(t)$ is the original signal of the human speaker, $E(t)$ and $M_{mic}(t)$ are impulse responses of the environment and microphone, $M'_{mic}(t)$, $S_{spk}(t)$ and $E'(t)$ are impulse responses in replay speech. From (1) and (2), the main difference between $S_{genuine}(t)$ and $S_{replay}(t)$ lies in three aspects, which are $M'_{mic}(t)$, $S_{spk}(t)$ and $E'(t)$. Firstly, the environmental factor $E'(t)$, if the environment is relatively quiet, this factor can be ignored. However, when the environment is noisy, this factor cannot be ignored. That means if $E'(t)$ and $E(t)$ differ greatly, and it will degrade the final performance of the system. As mentioned in [28], most systems will quickly fail when the noise increases. Secondly, $M'_{mic}(t)$ and $S_{spk}(t)$, these two factors are mainly related to the performance of the device. When high-quality recording equipment is used, these two factors will become smaller. Unlike the detection performance of replay attack, which is affected by the environment and equipment, the detection of DF speech is mainly affected by various algorithms. Different spoofing attack algorithms were used to create DF speech. In this paper, most of the DF speeches were generated by the latest state-of-the-art methods of TTS and VC. The number of algorithms exceeds one hundred. With some algorithms, such as Tacotron 2 [29], DF speech can be easily created. The DF task is without the ASV system in the ASVspoof 2021 challenge. The main goal is to protect the reputation of a well-known personality or detect disinformation or fake news [30].

### B. Audio Features

Generally, the system of spoofed audio detection uses a front-end and back-end structure. The front end extracts audio features, and the back-end classifies them according to these features. The minimum requirements of the classifier are to judge whether it is spoofed or bona fide speech. We can use a binary classification for simplicity. Here we mainly discuss the feature extraction of the front end. Feature extraction is generally performed after speech pre-processing. The standard pre-processing methods are frame division, emphasis, etc. In the specific operation process, there may be other operations, such as intercepting or extending the length of speech, to ensure that the length of all speeches is the same. After pre-processing, the speech features can be extracted. Traditional methods such as Mel-frequency cepstral coefficients (MFCC), Filter Banks (FBank), Constant Q Cepstral Coefficients (CQCC), etc., are fixed filter banks. The Leaf of the variable filter bank proposed by Google is to adjust the filter to be variable [31]. Leaf has achieved good results in speech applications, such as speech scene classification. In [28], Qian *et al.* also proposed a deep learning method. The features obtained through this method are also directly applied to Spoofed Audio detection, and a relatively ideal effect has been achieved in the ASVspoof 2015 dataset. In the application of spoofed audio detection, both traditional features and deep features have shown good results. The advantages and disadvantages of these two feature extractions are also very obvious. Traditional features have better generality and have achieved good results in various tests [12], [32]. However, traditional features have not achieved the best results compared to deep features. Deep features, through some complex designs [33], [34], can achieve excellent results. This excellent performance can be extended to practice if it is not the case of overfitting on some occasions [21].

### C. Classifier

There are two types of back-end classifiers. Gaussian Mixture Model (GMM), Linear Discriminant Analysis (LDA) and Support Vector Machine (SVM) are the traditional classifiers. As mentioned in Section I, the traditional classifier cannot get the best result in the ASVspoof 2015, 2017, 2019, and 2021 challenges. The other type is the classifier based on a deep learning network, such as a convolutional neural network (CNN), recurrent neural network (RNN), deep neural network (DNN), etc. In recent years, for the detection of spoof speech, most of those who have achieved good results use deep learning networks, such as [16], [17], [33], [35]. But those deep learning-based classifiers [36] need more data than the traditional ones [37]–[39]. Especially without a good enough feature, it is prone to be overfitting.

## III. Proposed Method

In this section, we proposed a novel framework based on audio hybrid features and a self-attention mechanism for replay and DF audio detection. First, it is the overview of the proposed framework, and then we will describe our motivation, audio hybrid features extraction, Self-attention mechanism, and loss function in detail.

### A. Overview

The audio signal will be put into two parallel paths in our proposed framework. The first path, shown in the blue box in Fig. 2 (a), took the audio signal with pre-processing and got the deep features through CNN. The second path, which was shown in the golden box in Fig. 2 (a), took pre-emphasis, resize, framing, and Short-Time Fourier Transform (STFT) and got Mel-spectrogram features. Then with the concatenation max pooling layer, it will get the audio hybrid features. These hybrid features will be put into the Self-attention module, which is shown in Fig. 2 (b). After this module, there were some ResNet Blocks, which are shown in Fig. 2 (c). Finally, there was a liner layer, and we could get the binary output. Details of the proposed framework are shown in Fig. 2.

### B. Motivation

Parallel networks have shown promising results in audio forgery detection [8], [40], and neural speech enhancement [41]. And feature fusion also can be used to improve the performance in speech recognition [42]. The learnable front end is another idea for audio feature extraction [31], [43]. Such audio features have achieved better results than fixed parameter features in many audio tasks. Therefore, we hypothesize deep features or more than one type of feature can capture sufficient audio feature information. In this work, we hoped that the combination of deep features and Mel-spectrogram features could get better feature expression. The disadvantage is that the weighting coefficients of the two features are difficult to determine, and different parameters or even the entire model need to be modified for different forgery methods. Inspired by the transformer network in speech feature extraction [44], [45], we took the Self-attention module to deal with the weight issue among two features.

In summary, we adopted a deep feature extraction method combining Mel-spectrogram features by parallel paths and used self-attention to weigh the features. We used ResNet as the back-end classifier to get the final result.

### C. Hybrid Features Extraction

In our approach, hybrid features were extracted, composed of deep features and Mel-spectrogram features. As mentioned in [46], it is important to get discriminative frequency information. Mel-spectrogram features contain almost all that information. After necessary pre-processing, we can get Mel-spectrogram features. The specific method is as follows:

1) Pre-emphasis.

$$Y(t) = S(t) + \alpha S(t-1) \tag{3}$$

where $\alpha = 0.97$, $S(t)$ and $S(t-1)$ is the original audio signal, and $Y(t)$ is the result of pre-emphasis. This step aims to enhance the energy of the high-frequency bands. 2) Unify the length of the audio signal, cut off the too-long voice signal, and add the too-short signal (Padding method). 3) Framing with Hamming window. The frame size is 32 milliseconds (ms), and the step size is 16 ms. That will be a 50% overlap. 4) With STFT and Mel-filter banks, we can get Mel-spectrogram. The wave plot and Mel-spectrogram are shown in Fig. 3 and Fig. 4. 5) After Batch Normalization, that will be the Mel-spectrogram features.

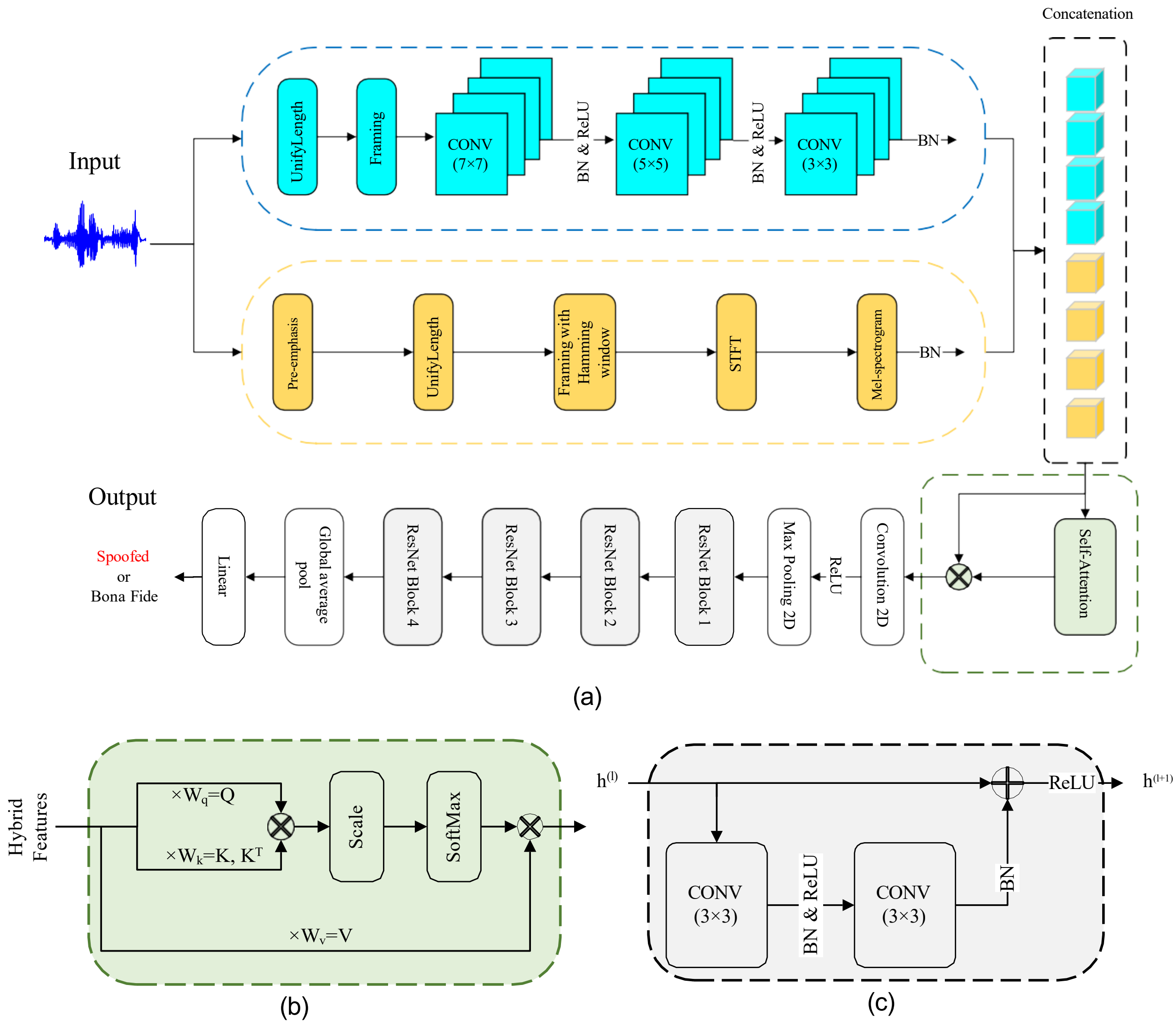

Fig. 2. (a)The proposed framework. (b)Self-attention module. (c) ResNet Block. BN means Batch Normalization and ReLu means ReLU Activation Function. ⊗ denotes Hadamard product. ⊕ denotes element-wise addition.

Deep features were extracted by CNN. The main steps were as follows: 1) Audio signals need to be unified the length before extracting deep features. It is the same as the second step of the Mel-spectrogram features. 2) Audio signals will be divided into frames, then we can obtain a short-term stable signal representation. Similar to the framing of the standard features, frame size, and frameshift will be fixed. We set the same size with the third step of the Mel-spectrogram features without Hamming window. We randomly selected one frame, about 32 ms, shown in Fig. 5. The amplitude of this frame could be considered constant. With the overlap of a half frame, no information will be dropped. 3) After the above steps, audio signals are still high-dimensional. Then we took these signals as input to CNN with a 7×7 kernel. We took these kernels as filters. 4) without stride, we get the exact size output as input. After Batch Normalization and an activation function of ReLu, it is another layer of CNN with a 5×5 kernel. Then repeat this step with a 3×3 kernel. 5) After Batch Normalization, that will be the Deep features. Concatenate Deep features and Mel-spectrogram features without any weight. We will get hybrid features.

### D. Self-Attention Mechanism

The architecture named Transformer [47] has been successfully applied in various fields. In many computer vision tasks, self-attention played a more and more important role. In [48], [49], they believed that the module could get better deep feature representation with the attention mechanisms by computing a weighted sum of features. In [50], [51], they got competitive results on some challenging computer vision tasks by self-attention-based methods. Moreover, in [52], they made the self-attention mechanism more explainable. Inspired by the success of the vision domain with self-attention, we supposed it could be transferred to the task of spoof audio detection. In fact, [44], [45], [53] used the Transformer-based method to get better results. In [54], they try to use self-attention as the critical function to get a better detection capacity for partially fake audio.

In this paper, with a self-attention mechanism, we try to find the autocorrelation for the feature map and get the most relevant features. Self-attention quite uses as a weighted Mask, that is, a matrix of weighting coefficients. Since the feature map is a superposition of two types of features, this feature map is redundant in the application of audio spoof detection.

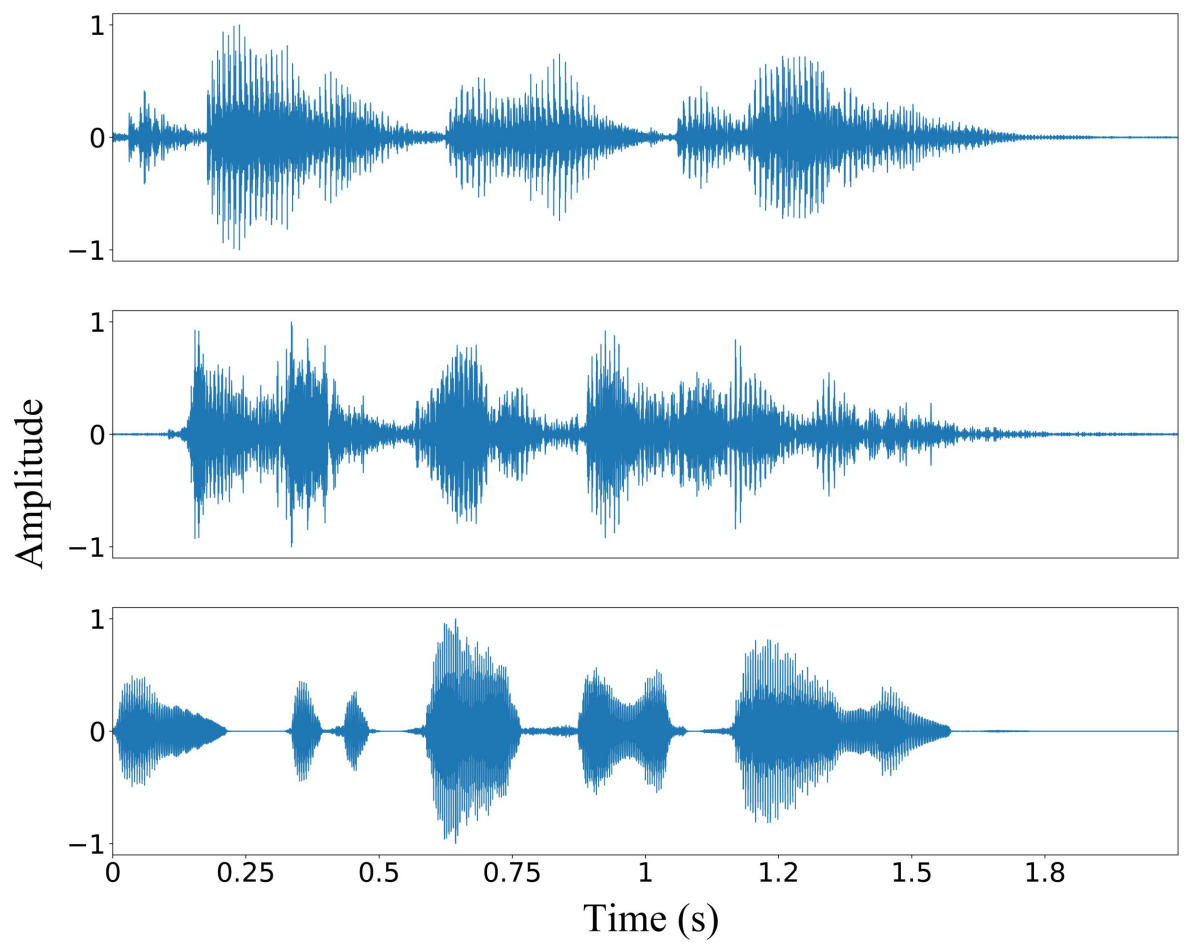

Fig. 3. The wave plot from up to down: bona fide, replay, and Deep-Fake audio clips.

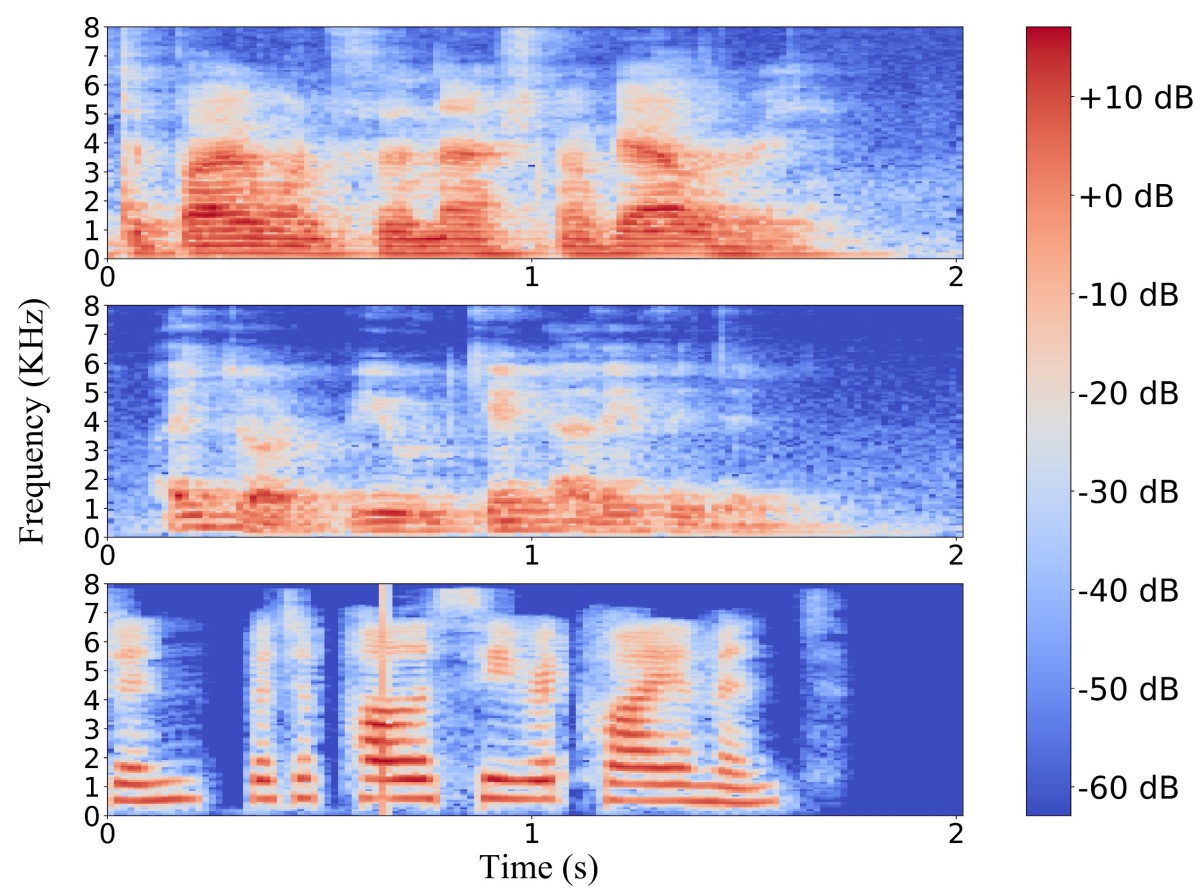

Fig. 4. The Mel-spectrum from up to down: bona fide, replay, and Deep-Fake audio clips.

We took $S = [s_1, s_2, s_3, ...s_T]$ as the input of deep features and set $T = 126$. After $7 \times 7$, $5 \times 5$, and $3 \times 3$ CNNs, we got the same size as input:

$$f(S) = [z_1, z_2, z_3, ...z_T] \tag{4}$$

where $f$ is the CNNs with the BN and ReLu in the middle, shown in the blue box in Fig. 2 (a). The Sampling rate of all audio files in our experiment is 16 kHz. It means $f(S) \in R(d_{cnn}, T)$, where $d_{cnn}$ is 512.

Let $M = [m_1, m_2, m_3, ...m_T]$ denote the $T$ frames of Mel-spectrogram features. In this paper, we took the number of 128 for Mel frequency banks. Here, $M \in R(d_{mel}, T)$, and $d_{mel}$ is 128.

Then in the concatenation layer, we vertically concatenate $f(S)$ and $M$. The final results were the hybrid features:

$$H = [h_1, h_2, h_3, ...h_T] \tag{5}$$

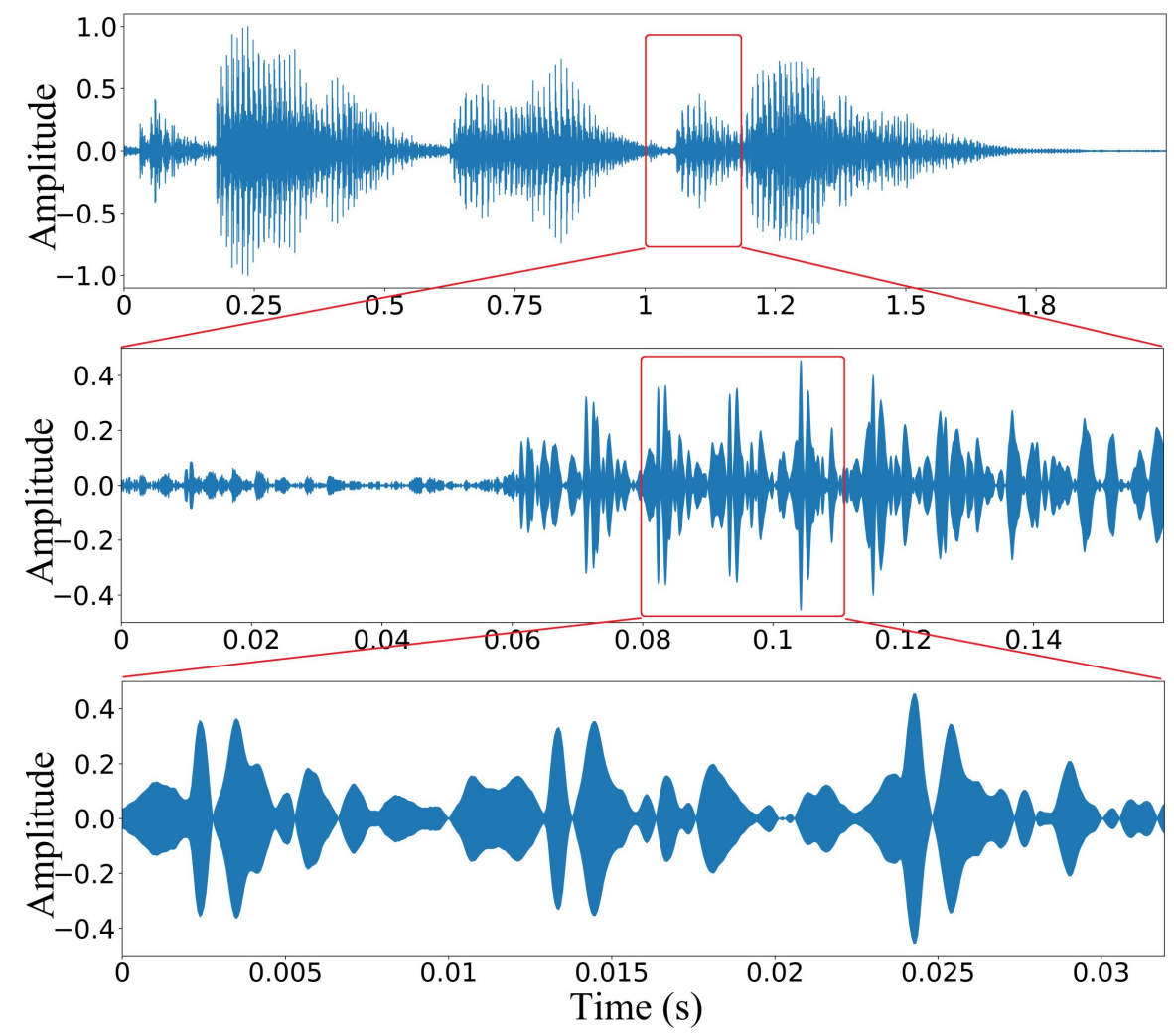

Fig. 5. The length of the top is 2 seconds, the middle is about 160 ms and the bottom is about 32 ms; that is the frame size.

where $H \in R(d, T)$, and $d$ is 640.

The self-attention layer is one of the key modules in Transformer [47]. It can be described as a mapping query and a set of key-value pairs to an output. In this paper, the queries, keys, and values are the set of collaborative item representations $H$. In the self-attention layer, which was described in Fig. 2 (c), we took these steps:

$$Q = H \times W_q \tag{6}$$

$$K = H \times W_k \tag{7}$$

$$V = H \times W_v \tag{8}$$

where $W_q, W_k, W_v \in R(d, d)$ and $Q, K, V \in R(d, T)$.

Then we took matrix transpose for $K$, and that is $K^T$. Finally, we can get the output:

$$O_{att}(Q, K, V) = softmax\left(\frac{QK^T}{\sqrt{T}}\right) V \tag{9}$$

$\sqrt{T}$ is the scale factor used to avoid that softmax function will get into extreme regions due to the large dot products growing large in magnitude. $O_{att}$ is the final output in this layer. It can be considered that this is the result of hybrid features $H$ after weighting.

### *E. Loss Function*

Our proposed neural networks can be trained as a binary classifier by the Sigmoid activation function and the binary cross entropy (BCE). It means that we can select the BCE Loss function for simplicity. As mentioned in Section I, we need to consider the model's generalization. Or it will suffer from performance degradation when there are some ”unknown” or brand-new DF or replay speech in detection tasks. Especially, there were more than one hundred algorithms that were used to generate DF speech. Our module should learn a robust representation across different domains. As suggested in [18], it should build a loss function to fit these tasks after mapping audio features to the embedding space. The loss function for training our proposed module should consider the feature extraction step, attention block, and the final classification losses. We use the unified Loss Function of the front and back ends. This allows us to optimize the integration system in the whole process of training.

$$L_{att} = -\sum_{i=1}^{T} O^i_{att} \log \hat{O}^i_{att} + \left(1 - O^i_{att}\right) \log \left(1 - \hat{O}^i_{att}\right) \quad (10)$$

$$L_{fin} = -\sum_{n=1}^{N} O^n \log \hat{O}^n + (1 - O^n) \log \left(1 - \hat{O}^n\right) \quad (11)$$

$$Loss = L_{att} + \lambda L_{fin} \quad (12)$$

where $L_{att}$ is the loss of the attention and feature extraction step and $L_{fin}$ is the loss of the final classification. Then we can get the loss of the system with the hyperparameter $\lambda$.

## IV. Experiments and Results

This section describes all experiments, including datasets, metrics, baseline system, experiments setup, and results.

### A. Datasets

Our experiments used two datasets: the ASVspoof 2019 and the ASVspoof 2021. All audio files are free lossless audio codec (FLAC) format. The sampled rate of all audio data is 16 kHz, and audio is stored in 16-bit. We took the ASVspoof 2019 for training and evaluated it in the ASVspoof 2021 dataset.

1) The ASVspoof 2019 dataset: There are three primary forms of spoofing attack, TTS, VC, and replayed speech, in this dataset. It contains two parts and is an independent evaluation algorithm. Among them, the logical access (LA) scenario part includes TTS and VC. We select the LA scenario part for module training in the DF task. The physical access (PA) scenario only has replayed speech with a high-quality device. Table I describes the ASVspoof 2019 dataset.

TABLE I
Description of the ASVspoof2019 dataset [4]

| | | #Utterances | | | |
|---|---|---|---|---|---|
| | | Logical Access | | Physical Access | |
| Subset | #Speakers | Bona fide | Spoof | Bona fide | Spoof |
| Training | 20 | 2580 | 22800 | 5400 | 22800 |
| Development | 10 | 2548 | 22296 | 5400 | 24300 |
| Evaluation | 48 | 7355 | 63882 | 18090 | 116640 |
| Total | 78 | 12483 | 108978 | 28890 | 163740 |

Especially, there is another subset in the ASVspoof 2019 dataset. It is a real replay subset that considers attacks in the real scenario. The real replay data contains an additive ambient noise [21]. Table II describes the real replay subset in the ASVspoof 2019 dataset.

2) The ASVspoof 2021 dataset: In this dataset, there are three parts: LA, PA, and DF. We selected PA and DF in this paper. PA task was more complicated than the previous ASVspoof challenge. It was in real, variable physical spaces with reverberation and additive noise to detect attacks [5]. We need to train our module in the ASVspoof 2019 dataset PA scenario. Many VC and TTS generate utterances in the DF scenario with different vocoders for transcoding. It is similar to the audio in social media. Table III is a description of the ASVspoof 2021 dataset without the LA scenario.

TABLE II
Description of the real replay subset in the ASVspoof 2019 dataset

| | | # Utterances | |
|---|---|---|---|
| | | Physical Access | |
| Subset | # Speakers | Bona fide | Spoof |
| Real Replay | 26 | 540 | 2,160 |

TABLE III
Description of the ASVspoof2021 dataset [30] (without LA)

| | #Utterances | | | |
|---|---|---|---|---|
| | Physical Access | | Deep-Fake | |
| Subset | Bona fide | Spoof | Bona fide | Spoof |
| Progress | 14,472 | 72,576 | 5,768 | 53,557 |
| Evaluation | 94,068 | 627,264 | 14,869 | 519,059 |
| Total | 108,540 | 699,840 | 20,637 | 572,616 |

In the ASVspoof 2021 challenge, several hidden subsets exist in LA, PA, and DF scenarios. The first type of hidden subsets is composed of those audios with non-speech intervals deleted automatically. Another hidden subset is only in the PA scenario containing simulated replayed data.

### B. Metrics and Baseline Systems

1) EER: The Equal Error Rate is the primary evaluation metric for the ASVspoof 2015 and ASVspoof 2017 challenges. In the ASVspoof 2021 DF task, it is the unique evaluation metric. When a certain threshold $\theta$ is taken, we can get the False Rejection Rate (FRR) and False Acceptance Rate (FAR). We can define $FRR$ and $FAR$ as follows:

$$FRR(\theta) = \frac{\#\{\text{Bona fide trials with score} \leq \theta\}}{\#\{\text{Total Bona fide trials}\}} \quad (13)$$

$$FAR(\theta) = \frac{\#\{\text{Spoof trials with score} > \theta\}}{\#\{\text{Total Spoof trials}\}} \quad (14)$$

where the score is the output of the classifier. A high score denotes bona fide audio, and a low score denotes spoof audio. There is no need to fix a certain threshold $\theta$ to use this metric. We can take a $\theta_{eer}$ to satisfy the following formula:

$$EER = FRR(\theta_{eer}) = FAR(\theta_{eer}) \quad (15)$$

2) t-DCF: Except for the DF task, the minimum tandem detection cost function (t-DCF) is the primary evaluation metric for the ASVspoof 2019 and the ASVspoof 2021 challenge. This metric is mainly used to evaluate the performance of the integrated system at the decision-making level. For example, the method includes two independent systems, the ASV and Presentation Attack Detection systems (PAD). It is necessary to consider the costs of missing target users and falsely accepting zero-effort impostors or spoofing attacks [55]. EER cannot measure these indicators. Therefore, the organizers of the ASVspoof challenge adopted t-DCF. The minimum t-DCF can be defined as:

$$t - DCF_{min} = \min_{\theta} \left\{ \beta P_{fr}^{PAD}(\theta) + P_{fa}^{PAD}(\theta) \right\} \quad (16)$$

where $P_{fr}^{PAD}$ and $P_{fa}^{PAD}$ are the PAD system false rejected and false alarm rates at threshold $\theta$. Coefficients $\beta$ depend on the ASV system and application parameters (priors, costs). Tomi Kinnunen *et al.* introduces t-DCF in detail in [42]. In [55], specific configuration parameters for this metric in the ASVspoof 2019 challenge are described.

3) Baseline systems: There are four baseline systems in the ASVspoof 2021 challenge. Baseline 1 (B1) uses constant-Q cepstral coefficients (CQCC) with GMM [56]. Baseline 2 (B2) uses linear frequency cepstral coefficients (LFCC) with GMM [57]. Baseline 3 (B3) uses LFCC with a light convolutional neural network (LCNN) [58]. Baseline 4 (B4) uses RawNet2 [59]. B1 almost got perfect results on the ASVspoof 2015, and BTAS2016 datasets [34]. With the help of deep learning architecture, both B3 and B4 have achieved excellent results in some dataset experiments. It is worth noting that B4 is also an end-to-end detection system.

### C. Settings

In our experiments, the audio signal is simply pre-processed. After pre-emphasis with $\alpha$ = 0.97, we cut the length to 2 seconds since the length of speech is different. If the size is not enough, we will use the copy method to extend it to this length. Then we use the hamming window to divide the speech into frames. The frame length is 32ms, and the step length is 16ms, which means that 16ms of adjacent frames are overlapped. It is the same configuration in the deep features extraction path but with no pre-emphasis and hamming window. The initial weights of training are randomly generated. The mini-batch size we used is 16 to 64, and the learning rate of 1e-4 to 2e-2. In the initial stage of training, we use only the training part of the data in the ASVspoof 2019 dataset. Then we verify our method in the development section. At this time, adjust the hyperparameters to get better results. When the evaluation part of the data is used in the subsequent test, the hyperparameters will be fixed and no longer adjusted. Details of Self-attention and hybrid features-based architecture are shown in Table IV.

TABLE IV
DETAILS OF SELF-ATTENTION AND HYBRID FEATURES-BASED ARCHITECTURE

| Layers | Output size | Layer config |
|---|---|---|
| Convolution 1 | 512 × 126 | 7 × 7 |
| Convolution 2 | 512 × 126 | 5 × 5 |
| Convolution 3 | 512 × 126 | 3 × 3 |
| Pooling(concatenation) | 640 × 126 | - |
| Convolution | 640 × 126 | 7 × 7 |
| Max Pooling | 320 × 63 | 3 × 3 |
| ResNet block 1 | 320 × 63 | 3 × 3<br>3 × 3 × 2 |
| ResNet block 2 | 160 × 31 | 3 × 3<br>3 × 3 × 2 |
| ResNet block 3 | 80 × 15 | 3 × 3<br>3 × 3 × 2 |
| ResNet block 4 | 40 × 7 | 3 × 3<br>3 × 3 × 2 |
| Classification Layer | 1 × 1 | Global average pool, Linear |

### D. Results

In the ASVspoof 2019 LA scenario experiments, our proposed system outperforms baseline systems B1 and B2. But considering with min t-DCF metric, it did not get the best performance. The results of the ASVspoof 2019 LA scenario are shown in Table V.

TABLE V
RESULTS OF THE ASVSPOOF 2019 LA SCENARIO

| Individual system | min t-DCF | EER(%) |
|---|---|---|
| B1 | 0.2839 | 9.57 |
| B2 | 0.2605 | 8.09 |
| FFT-LCNN [60] | **0.1028** | 4.53 |
| **Self-attention and Hybrid Features(Proposed)** | 0.1039 | **4.49** |

In the ASVspoof 2019 PA scenario experiments, real replay attacks were included. Unlike well-designed simulated data, real replay data is closer to real application scenarios. The results show that additive noise has a non-negligible impact on system performance. The performance of each detection system degrades with the real replay data. Notably, the CQT-LCNN system achieves the best performance in the evaluation stage [60], but the EER increases to 25.14% with the real replay data. Our proposed system achieves the best performance with a slight drop in performance with the real replay data. All results of the ASVspoof 2019 PA scenario with the real replay subset are shown in Table VI.

TABLE VI
RESULTS OF THE ASVSPOOF2019 PA SCENARIO AND REAL REPLAY DATA

| Individual system | Eval | | Real Replay | |
|---|---|---|---|---|
| | min t-DCF | EER(%) | min t-DCF | EER(%) |
| B1 | 0.3476 | 11.04 | 0.3855 | 12.73 |
| B2 | 0.3481 | 13.54 | 0.6681 | 29.44 |
| CQT-LCNN(1) | **0.1626** | **1.23** | 0.7224 | 25.14 |
| **Self-attention and Hybrid Features (Proposed)** | 0.2157 | 6.37 | **0.3025** | **8.95** |

(1) Proposed in [60], results were obtained by us

The ASVspoof 2021 PA scenario experiment, the results of EER and t-DCF were considered. The verification results on the Development dataset are obtained after training on the Training dataset in the ASVspoof 2019 dataset. The final results on the Evaluation subset are obtained after verifying and adjusting the hyperparameters on the progress subset in the ASVspoof 2021 dataset. The minimum t-DCF is calculated with the ASV system provided by the ASVspoof organizer. For comparison, we took the results of six individual systems, including four baseline systems, the T07 (It is the ID of a participant in the challenge) system which was the best result of the ASVspoof 2021 challenge in the PA scenario, and a device features based system. The results are shown in Table VII. Our proposed system gets the best result and improves the EER by 74.60% more than B1, which is the best baseline system in the PA scenario.

TABLE VII
RESULTS OF THE ASVSPOOF2021 PA SCENARIO

| Individual system | t-DCF | EER(%) |
|---|---|---|
| Device Features Based [40] | 0.9503 | 35.61 |
| T07 [5] | 0.6824 | 24.25 |
| B1 | 0.9434 | 38.07 |
| B2 | 0.9724 | 39.54 |
| B3 | 0.9958 | 44.77 |
| B4 | 0.9997 | 48.60 |
| **Self-attention and Hybrid Features(Proposed)** | **0.4376** | **9.67** |

In the ASVspoof2021 DF task experiment, a similar procedure was also adopted. The difference is without t-DCF, take EER for comparison in all individual systems. This part of the dataset is closer to the real social media scenario and contains a certain amount of noise. We took the results of four baseline systems and the T23 (It is the ID of a participant in the challenge) system which was the best result of the ASVspoof 2021 challenge in the DF task for comparison. The results are shown in Table VIII. Our proposed system gets the best result and improves the EER by 60.05% more than B4, the best baseline system in the DF task.

TABLE VIII
RESULTS OF THE ASVSPOOF 2021 CHALLENGE DF TASK

| Individual system | EER(%) |
|---|---|
| T23 [5] | 15.64 |
| B4 | 22.38 |
| B3 | 23.48 |
| B2 | 25.25 |
| B1 | 25.56 |
| **Self-attention and Hybrid Features(Proposed)** | **8.94** |

# V. DISCUSSIONS

## A. Feature or Classifier?

In the task of detecting spoof audio, we need to build a better feature or design a more robust classifier. But which one is more important? Feature or Classifier? We took more tests in the PA and DF tasks. At first, we took one type of feature. The results are shown in Table IX. Deep Features got the best result in PA and DF tasks. The other system with one type of feature cannot get a good enough result in the PA task. But the gap is modest in the DF task. The MFCC-based system performed better than the CQCC-based and LFCC-based in the DF task.

TABLE IX
RESULTS OF THE ONE TYPE OF FEATURE

| Individual system | PA EER(%) | DF EER(%) |
|---|---|---|
| Self-attention + MFCC | 36.84 | 18.57 |
| Self-attention + CQCC | 28.31 | 20.34 |
| Self-attention + LFCC | 31.56 | 19.87 |
| **Self-attention + Deep Features** | **19.73** | **16.86** |

We took some other tests for the classifier. By replacing the classifier with or without Self-attention in our proposed method, we took this experiment in PA and DF tasks. The results are shown in Table X. Without Self-attention, all systems suffer performance degradation. It was interesting that RawNet2 got the best result and DenseNet [61] almost got the worst result with or without Self-attention in the DF task. We supposed that the reason for the performance degradation of DenseNet is the overfitting of the system.

TABLE X
RESULTS OF HYBRID FEATURE WITH OR WITHOUT SELF-ATTENTION

| Individual system | PA EER(%) | DF EER(%) |
|---|---|---|
| Hybrid Feature +GMM | 36.45 | 31.14 |
| Hybrid Feature +RawNet2 | 39.52 | 28.43 |
| Hybrid Feature +DenseNet | 31.26 | 33.02 |
| Self-attention + Hybrid Feature +GMM | 20.45 | 21.28 |
| Self-attention + Hybrid Feature +RawNet2 | 18.42 | **17.58** |
| Self-attention + Hybrid Feature +DenseNet | **9.89** | 31.41 |

We hardly tell that feature or classifier which is more important in the detection task from the results in Tables IX and X. However, we can find that Self-attention plays an essential role in these systems, and the deep-learning-based classifier can get convincing performance, but be careful to avoid the issue of overfitting. These results also verified our previous hypothesis in Section III that deep features or more than one type of feature can get sufficient distinguishing information in these detection tasks. Therefore, we believe that in the application field of this paper, features, and classifiers have equal weight.

## B. Frame Size Matters?

The Frame size selected in this paper is 32 ms. Experiments show that the size change will significantly influence the final result, as shown in Fig. 6. We took these experiments with different features in PA and DF tasks. The system with the Hybrid Feature proposed by us can get more robust results when the frame size changes, shown in Fig. 6 (a). From 20 ms to 45 ms, the system's performance changes little. The same situation also occurs in deep feature-based systems, shown in Fig. 6 (b). But the system consistently performs over a wider frame size range, 25ms to 60ms in PA and DF tasks. The performance of the MFCC feature-based system shows that PA and DF tasks should select different frame sizes for better results, as shown in Fig. 6 (c). For the PA task, the system should select a frame size of 20 ms and 40 ms for the DF task. In the DF task, the system, with the CQCC feature, can select a frame size of 30 ms to 55 ms, shown in Fig. 6 (d). But it can select a frame size of 30ms to 40ms. Otherwise, the performance will decrease.

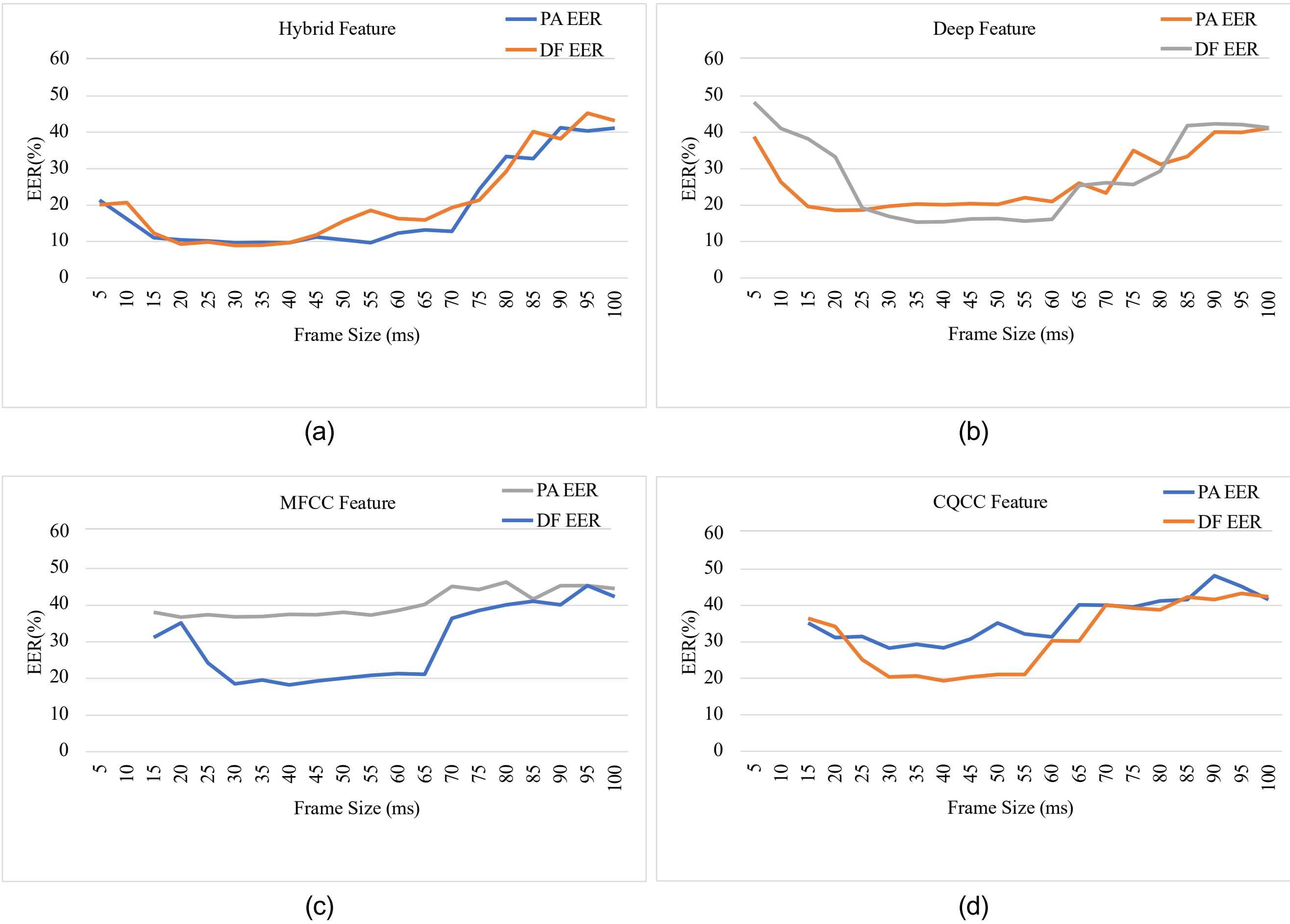

Fig. 6. The performance of each system when the frame size changed in PA and DF tasks. (a) Our proposed hybrid feature system. (b) Deep feature-based system. (c) MFCC feature-based system. (d) CQCC feature-based system.

Above all, the frame size explicitly impacts the system's performance. At the same time, in the case of different frame sizes, our proposed system can still perform well. This also proves that hybrid features can better extract the essential features of the audio signal from another perspective.

### *C. About Generalization*

In the test of the ASVspoof2021 dataset, we further add the hidden subset to explore the system's generalization. In the LA scenario, the hidden subset data automatically removes non-speech segments. This means that the system cannot extract features from speech intervals. Fig. 7 (a) shows that without non-speech segments system's EER increases by nearly 20%. T23 and T07 are the best-performing detection systems in the ASVspoof 2021 Challenge [21]. The change in EER achieved by our proposed system is the smallest, but 19 percentage points also increase the EER. This result also shows that in the detection of the LA scenario, the detection system seems to over-rely on the features of non-speech segments. In the task of DF, the increase of hidden subsets does not seem to impact the system's performance much. Detailed results are shown in Fig. 7 (b). Baseline system B3 has the largest EER increase of about 8% without non-speech. The EER changes of several other detection systems are all within the 5% range. The performance of our proposed system is hardly affected by the hidden subset, with less than a 1% change in EER. This also proves that the generality of our proposed Self-Attention and Hybrid Features system is better than other audio spoof detection systems.

In the PA scenario, the first hidden subset includes data with or without non-speech segments. The second hidden subset contains simulated replay data. Although it is mentioned in [21] that simulated data significantly impacts minimum t-DCF when only EER is considered, the system's performance is nearly unaffected by non-speech and simulated data. Detailed results are shown in Fig. 8. The T07 system, one of the systems that performed best in the ASVspoof 2021 challenge, had little or no change in EER in the two hidden subsets. The proposed Self-Attention and Hybrid Features system also shows good generalization, and EER is not affected by hidden subset data. In the DF task, evaluation utterance is a different source, with additive noise, encoding, compression, and transmission.

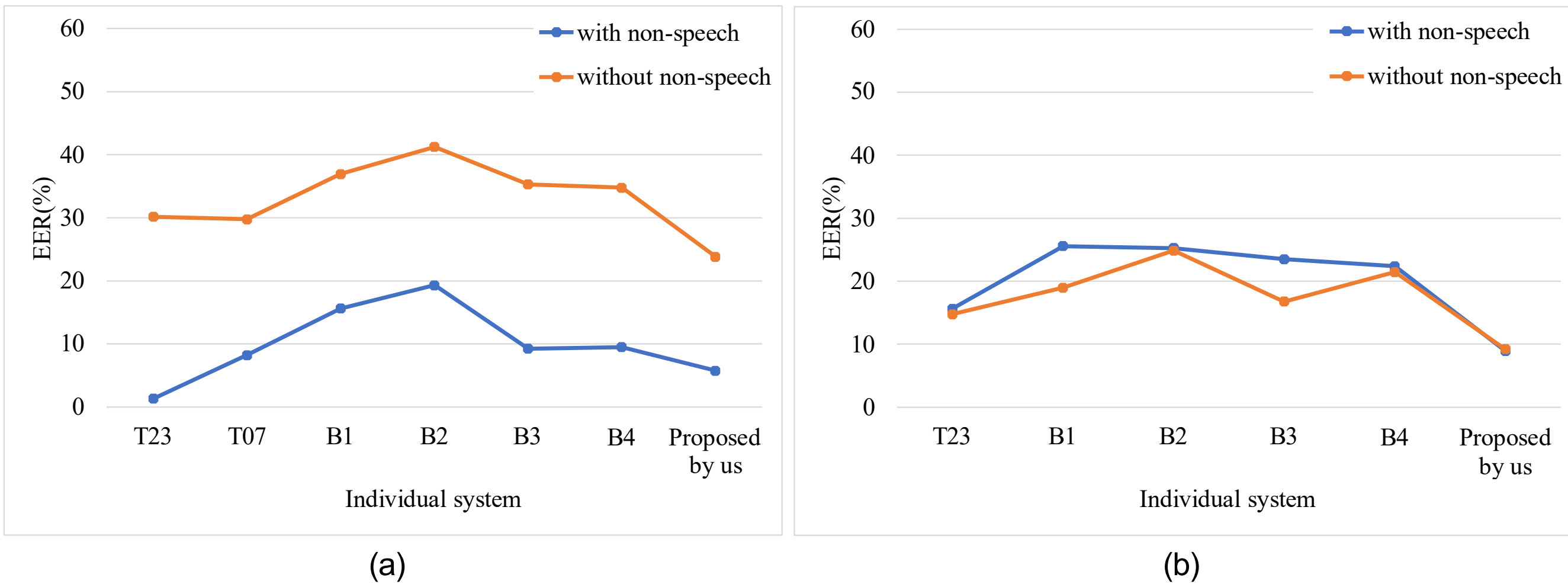

Fig. 7. Results of the detection systems in the ASVspoof 2021 LA and DF tasks with or without non-speech segments. (a) Evaluation results in the ASVspoof 2021 LA task with or without non-speech segments. (b) Evaluation results in the ASVspoof 2021 DF task with or without non-speech segments.

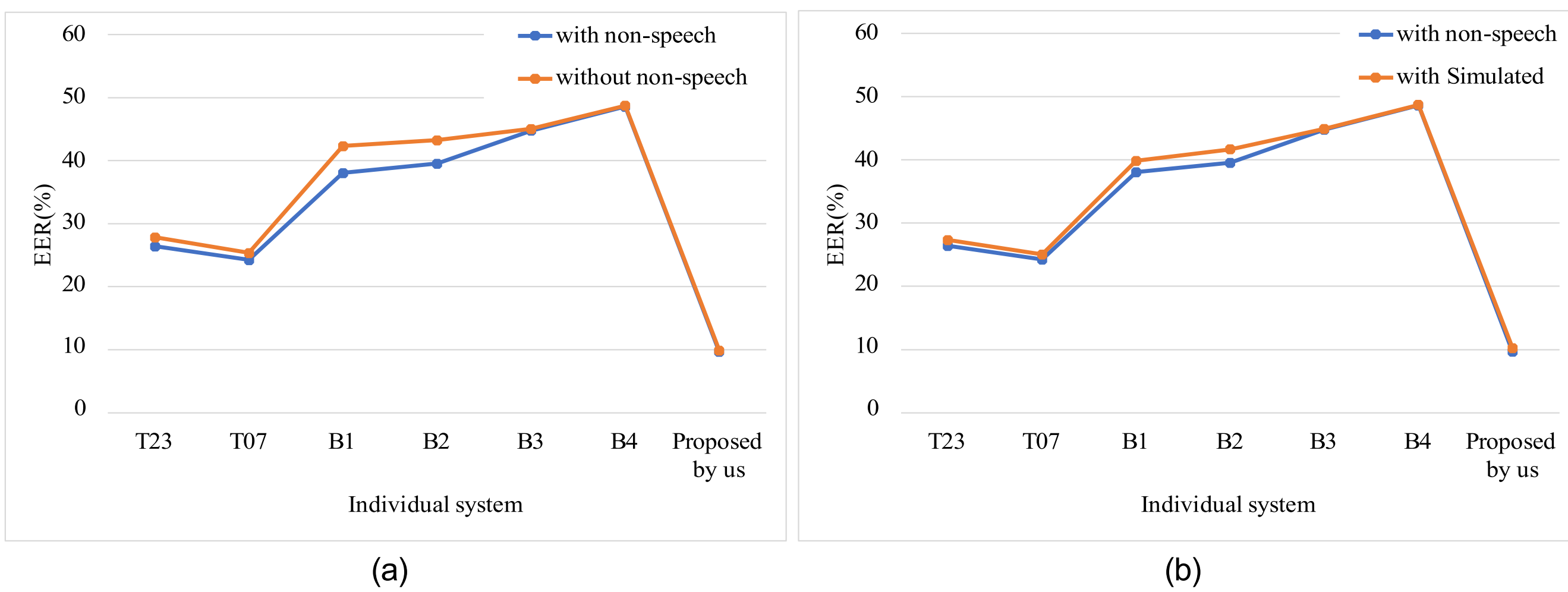

Fig. 8. Results of the detection systems in the ASVspoof 2021 PA tasks with or without non-speech segments and simulated replay data. (a) Evaluation results in the ASVspoof 2021 PA task with or without non-speech segments. (b) Evaluation results in the ASVspoof 2021 PA task with or without simulated replay data.

Most of them cannot be seen in the training stage. It can be explained that our proposed model is more general and can be robust in noise. Deep learning or hybrid features may become the best alternative to traditional ones. This paper optimizes system performance from two aspects: classifier and feature extraction. It can be seen from the results of the experiment that a single improvement (classifier or feature extraction) can improve system performance.

Self-attention increases the complexity of the system. Can it be replaced by directly adding several layers of a network? In this model, the primary function of Self-attention is to weigh several features to focus more on some more general essential features. Therefore, directly increasing the network layers is more challenging to achieve this function.

## VI. Conclusion

In this paper, CNN is used to extract the deep features from an audio signal, and Mel-spectrogram features are extracted at the same time. Then deep features and Mel-spectrogram features are concatenated into hybrid features. The Self-attention mechanism is used to weigh the hybrid features. Finally, ResNet is used for classification. The experimental results show that our method has achieved very good results in the DF and PA scenarios in the ASVspoof 2021 dataset. It offers the best performance in the DF and PA subsets, indicating that the method is more robust to noise. Experimental results reveal that the hybrid features, compared with conventional features, can cover more details of an utterance. In addition, the tests on the hidden subsets on the ASVspoof 2021 show that our framework provides good generalization. Without the Self-attention module, the system suffers performance decreasing. This highlights the critical role of the Self-attention module. The classifier framework used in the model in this paper is ResNet. It can be considered that this classifier still has excellent potential for improvement, such as replacing it with more layers or more complex DNN when resources permit. Therefore, we believe that the fusion of multiple features may be one of the development trends in this field.